# Nucleation of amyloid fibrils

Dimo Kashchiev

*Institute of Physical Chemistry, Bulgarian Academy of Sciences, ul. Acad. G. Bonchev 11, Sofia 1113, Bulgaria*

Stefan Auer [a)]

*Centre for Molecular Nanoscience, University of Leeds, Leeds, LS2 9JT, UK*

**ABSTRACT**

We consider nucleation of amyloid fibrils in the case when the process occurs by the mechanism of direct polymerization of practically fully extended protein segments, i.e. β-strands, into β-sheets. Applying the classical nucleation theory, we derive a general expression for the work to form a nanosized amyloid fibril (protofilament) constituted of successively layered β-sheets. Analysis of this expression reveals that with increasing its size, the fibril transforms from one-dimensional into two-dimensional aggregate in order to preserve the equilibrium shape corresponding to minimal formation work. We determine the size of the fibril nucleus, the fibril nucleation work and the fibril nucleation rate as explicit functions of the concentration and temperature of the protein solution. The results obtained are applicable to homogeneous nucleation which occurs when the solution is sufficiently pure and/or strongly supersaturated.

**PACS:** 64.60.Q-; 82.60.Nh; 87.14.E-; 87.15.nr; 87.19.xr

---

[a)]Author to whom correspondence should be addressed. Electronic mail: S.Auer@leeds.ac.uk

# I. INTRODUCTION

Understanding the nucleation mechanism by which proteins assemble into highly ordered structures known as amyloid fibrils is a much-studied problem because of its implications for human health and nanotechnology. At present there are about 25 different disorders categorized as amyloid diseases including Alzheimer's and Parkinson's disease,[1] and the application of peptide building blocks in bio-sensors, tissue engineering, and antibacterial agents has recently been demonstrated.[2] Structural studies[1,3] have shown that amyloid fibrils formed by different proteins are composed of protofilaments that are wound together to form higher-order fibrillar structures. The protofilaments themselves are composed of several β-sheet layers that share a common characteristic cross-β structure. The application of electron and X-ray diffraction and of solid state NMR spectroscopy to microcrystals of various short peptide fragments and to amyloid fibrils also gives evidence of this structure.[4-8]

It is now well established that fibrillar protein aggregates form through a nucleation mechanism.[9-33] Their formation kinetics is characterized by an initial lag time during which no aggregates are detected and a maximal rate of the overall aggregation process. Time-resolved optical experiments that measure the fluorescence signal arising from dye molecules such as thioflavin-T bound to the protein aggregates enable determination of the lag time and the maximal aggregation rate. Interestingly, the product of these two quantities is nearly the same for a wide range of aggregation conditions and protein systems.[34,35] So far, a considerable effort has been devoted to understanding how the amino acid sequence of proteins and the experimental conditions affect the kinetics of amyloid fibril formation.[30,32,33,36-44]

Experiments that investigate the physiochemical properties of the natural amino acids (such as β-propensity, hydrophobicity, aromatic content and charge) have been used to substantiate phenomenological models able to predict changes in the aggregation rate upon mutation as well as to predict amino acid sequences of proteins, so-called hot spots, that are likely to belong to the fibril core.[45-50] Although both the experimental studies and the theoretical models show that the kinetic parameters of aggregation depend strongly on the specificity of the amino acid sequence of the protein, it may be expected that this specificity is a particular expression of a common fibril nucleation/growth mechanism which could be treated in the framework of existing general theories of nucleation and growth of new phases.

Treating protein aggregation as a nucleation-mediated process is necessarily based on the concept that the process is reversible. This is so, because nucleation can only occur in a metastable protein solution, and the solution metastability can only be defined by means of the protein equilibrium (or saturation) concentration at which the bulk protein phase neither grows nor dissolves in the solution. Below this concentration (known also as solubility) the solution is undersaturated so that nucleation and growth are impossible, and existing protein aggregates dissolve. Both experiments (e.g., Refs. 18,30,44,51-54) and computer simulations (e.g., Refs. 31,55) give evidence for the existence of equilibrium peptide concentration and for the dissolution of amyloid fibrils in sufficiently dilute solutions. Knowing the equilibrium concentration of monomeric peptide allows determination of the supersaturation for fibril nucleation, and the fibril dissolution demonstrates that fibril formation is not an irreversible process.

In this article, our objective is to apply the classical nucleation theory (CNT) (e.g., Refs. 56,57) and the recently proposed corrected CNT (CCNT)[58] for describing the nucleation of amyloid fibrils by the mechanism of direct polymerization of practically fully extended protein segments, i.e. β-strands, into β-sheets. This mechanism is operative under conditions when no fibril precursors such as droplet-like peptide aggregates appear in the solution as a first step in the fibril nucleation process. By assuming that the nanosized amyloid fibril (protofilament) is built up of successively layered β-sheets, we derive a general expression for the work to form such a fibril. This expression is used for determination of the size of the fibril nucleus, the fibril nucleation work and the fibril nucleation rate. The results obtained are applicable to homogeneous nucleation which occurs when the protein solution is sufficiently pure and/or strongly supersaturated.

## II. PHYSICAL MODEL

The model we propose is based on structural and morphological studies of amyloid fibrils and microcrystals,[4,5,7,8,52,59] as well as on computer simulation results for amyloid fibril formation.[21-23,31,54] Our considerations thus pertain to nanosized amyloid fibrils (protofilaments) built up of successively layered β-sheets with fixed width and thickness.

Figure 1 schematizes such a fibril containing $i$ such β-sheets ($i \geq 1$), each β-sheet having $m$ practically fully extended rod-like peptides ($m \geq 1$) arranged parallel to each other and perpendicular to the fibril lengthening axis. The fibril is therefore prismatically shaped and contains a total of $n = im$ peptides ($n \geq 1$). The volume occupied by a peptide in the fibril is represented by that of right rectangular prism with cross-sectional area $a_0$ (m$^2$) and lateral areas $a$ (m$^2$) and $a_h$ (m$^2$) of its two different side surfaces (Fig. 1). These areas are given by $a_0 = d_h d$, $a = d_0 d$ and $a_h = d_0 d_h$, where $d_h$ (m) is the interpeptide distance in a β-sheet, $d$ (m) is the intersheet distance in the fibril, and $d_0$ (m) is the extended peptide length, i.e. the β-sheet width.

Since the fibril is with fixed width equal to that of the constituting β-sheets, it can be conceived as a two-dimensional (2D) phase in the $m,i$ plane. The fibril is thus congenial with a 2D rectangularly shaped crystallite and its nucleation, lengthening along the $m$ axis and thickening along the $i$ axis could be described in the scope of existing crystal nucleation and growth theories (e.g., Refs. 56,57). Important parameters in these theories are the specific surface energies of the different crystal faces. For the fibril in Fig. 1, we denote by $\sigma$ (J/m$^2$) and $\sigma_h$ (J/m$^2$) the specific surface energies of the faces perpendicular to the $m$ axis and the $i$ axis, respectively. The third fibril specific surface energy, $\sigma_0$ (J/m$^2$), is that of the fibril face in the $m,i$ plane. However, knowing this energy is not necessary for the present considerations, because they are restricted to the earliest stage of fibril formation when the fibrils are so small that they change their size solely along the $m$ and $i$ axes. As they are then 2D formations, $\sigma_0$ participates only implicitly, via the supersaturation $\Delta\mu$ (J), in the description of their nucleation (see Section III).

By definition, the specific surface energy is equal to the work done on creating a unit area of the respective face by cutting the fibril along a plane parallel to the face and separating the resulting two half-fibrils far enough from each other. For that reason, the $\sigma$ and $\sigma_h$ values are largely determined by the strength of the interpeptide bonds within the fibril. Whereas along the $m$ axis the peptide bonding is dominated by the hydrogen bonds,[60] along the $i$ axis the peptides (and, hence, the β-sheets) are bound by much weaker bonds such as those due to the hydrophobic effect. Therefore, since to a first approximation the surface energy is proportional to the bond energy, the amyloid fibrils are characterized by the important inequality $\sigma_h << \sigma$. For example, we shall have $\sigma_h = \sigma/10$ or $\sigma/20$ for fibrils with bond energy (per peptide) between two nearest-neighbor β-sheets that is about 10 or 20 times lower than the hydrogen bond energy between nearest-neighbor peptides in a β-sheet.

## III. WORK OF FIBRIL FORMATION

At a given absolute temperature $T$ (K) of the solution there exists a concentration $C_e$ ($m^{-3}$) of peptide monomers, called the equilibrium concentration or solubility, at which a macroscopically large fibril neither grows nor dissolves. The solution is then saturated and fibril nucleation is impossible. When the actual concentration $C_1$ ($m^{-3}$) of peptide monomers in the solution is lower or higher than $C_e$, the solution is undersaturated or supersaturated, respectively. While in the former case the system is in a thermodynamically stable state and fibril nucleation is again impossible, in the latter case the system is metastable and already able to nucleate fibrils that can grow up to macroscopic sizes. Using the physical model described above, we shall now determine the work $W_{i,m}$ (J) to form a nanosized amyloid fibril of thickness $i$, length $m$ and size $n = im$ ($n$ is the number of peptides constituting the fibril).

From nucleation theory (e.g., Refs. 56,57), the general expression for $W_{i,m}$ is of the form

$$W_{i,m} = -(\mu_s - \mu_f)im + \Phi_{i,m},$$ (1)

where $\mu_s$ (J) and $\mu_f$ (J) are, respectively, the chemical potentials of a peptide in the solution and in the bulk fibrillar phase of which the fibril is the precursor, and $\Phi_{i,m}$ (J) is the excess energy of the $i$,$m$-sized fibril. Physically, the first term on the right of Eq. (1) is the work gained from the assembling of $n$ peptides into an $n$-sized fibril, and the second term is largely the work done on creating the fibril/solution interface. Determining $\Phi_{i,m}$ is a hard physical problem, but with a relatively simple solution in the scope of CNT of crystals. According to this theory (e.g., Refs. 56,57), which we shall closely follow in the present study, $\Phi_{i,m}$ is merely the total surface energy of the $i$,$m$-sized fibril so that it can be written down as

$$\Phi_{i,m} = 2a_0\sigma_0 im + 2a\sigma i + 2a_h\sigma_h m.$$ (2)

Importantly, CNT neglects the possible dependence of $\sigma_0$, $\sigma$ and $\sigma_h$ on the fibril size. We note also that unlike the rigid crystallite, the fibril is a flexible formation which is why $\Phi_{i,m}$ from Eq. (2) should include additional *i,m*-dependent terms accounting for the fibril bending, twisting and other likely deformations. However, if these terms are incorporated into the summands in Eq. (2), the equation will formally remain unchanged, but with generalized and possibly *i,m*-dependent fibril specific surface energies. Assuming that such a dependence is sufficiently weak, in the spirit of CNT we shall hereafter restrict our analysis to *i,m*-independent $\sigma_0$, $\sigma$ and $\sigma_h$.

Combining Eqs. (1) and (2), we obtain

$$W_{i,m} = -im\Delta\mu + 2a\sigma i + 2a_h\sigma_h m , \qquad (3)$$

where the supersaturation $\Delta\mu \geq 0$, defined by

$$\Delta\mu = \mu_s - \mu_f - 2a_0\sigma_0 , \qquad (4)$$

is the driving force for nucleation of 2D fibrils and their further growth in length and thickness, i.e. along the *m* and *i* axes. From Eq. (3) we thus find that the dimensionless work $w_{i,m} \equiv W_{i,m}/kT$ to form an *i,m*-sized fibril is given by

$$w_{i,m} = -sim + 2\psi i + 2\psi_h m , \qquad (5)$$

where $s \equiv \Delta\mu / kT \geq 0$ is the dimensionless supersaturation, $k$ is the Boltzmann constant, and $\psi \equiv a\sigma / kT > 0$ and $\psi_h \equiv a_h\sigma_h / kT > 0$ are the dimensionless specific surface energies of the fibril faces perpendicular to the *m* axis and the *i* axis, respectively. Due to the much stronger hydrogen bonding in comparison with the hydrophobicity-mediated one, the inequality $\psi \gg \psi_h$ holds. As to *s*, Eqs. (A4) and (A9) in Appendix A show that it is given by $s = \ln(C_1 / C_e)$ or $s = L\Delta T / kT_eT$ for solutions in which the supersaturation is experimentally controlled by $C_1$ at a given $T$ or by $T$ at a given $C_1$ [ $\Delta T \equiv T_e - T$ (K) is the undercooling, $T_e$ (K) is the solution equilibrium temperature, and $L$ (J) is the latent heat or enthalpy (per peptide) of peptide aggregation].

Most generally, the fibril can thicken at fixed length and lengthen at fixed thickness which is why in Eq. (5) *i* and *m* can be treated as independent variables. This makes it possible to use Eq. (5) for determination of the work to form an *i*β-sheet of $n = im$ peptides (the *i*β-sheet is defined[55] as a fibril with a fixed number *i* of successively layered, equally long β-sheets in it). In particular, at $i = 1$ we have $n = m$ and from Eq. (5) we find that the formation of 1β-sheet of *n* peptides, i.e. of a single *n*-sized β-sheet, requires the work

$$w_{1,n} = -(s - 2\psi_h)n + 2\psi \,. \tag{6}$$

The straight lines 1β in Fig. 2 illustrate the $n$ dependence of the 1β-sheet formation work $w_{1,n}$ at scaled supersaturation $s/\psi_h = 1$, 2 or 3 (as indicated). The lines are drawn according to Eq. (6) with $\psi = 10\psi_h$.

Equation 6 is of special interest, for it reveals the conditions under which the 1β-sheet can or cannot form spontaneously. As seen from this equation and the lowest line 1β in Fig. 2, $w_{1,n}$ diminishes linearly with increasing the 1β-sheet size $n$ or, equivalently, length $m$ when $s > 2\psi_h$ which namely is the condition for spontaneous formation of 1β-sheets. On the contrary, at those low supersaturations that satisfy the condition $s < 2\psi_h$, more and more work has to be done on the 1β-sheet lengthening [$w_{1,n}$ from Eq. (6) then increases steadily with $n$; see the uppermost line 1β in Fig. 2]. Inasmuch as the fibril is the 1β-sheet's offspring, this implies that when $s < 2\psi_h$, the process of fibril formation is hampered by the existence of an energy barrier or, in other words, that it is nucleation-mediated. As visualized by the middle line 1β in Fig. 2, at the supersaturation

$$s_{1\beta} = 2\psi_h = 2a_h\sigma_h/kT \tag{7}$$

the work $w_{1,n}$ from Eq. (6) does not change with the 1β-sheet size $n$. Physically, this means that at $s_{1\beta}$ the 1β-sheet is in thermodynamic equilibrium (or coexistence) with the solution. According to Eq. (7), $s_{1\beta}$ scales with the fibril "hydrophobic" specific surface energy $\sigma_h$.

What is important about the 1β-sheet is that it cannot grow unlimitedly as one-dimensional (1D) formation, i.e. in length only. This is so, because when the 1β-sheet is constituted of too many peptides, its total surface energy may already be higher than that of a 2β-sheet of the same size $n$. As at fixed $n$ the fibril total surface energy has to be minimal, the most probable fibril shape will be the one that satisfies this thermodynamic requirement. This so-called equilibrium fibril shape is characterized by that value of the $i/m$ ratio which fulfils the condition for extremum of $w_{i,m}$ from Eq. (5). Since this condition is of the form $\partial w_{i,m}/\partial i = \partial w_{i,m}/\partial m = 0$ when $i$ and $m$ are treated as continuous variables, we readily find that the thickness $i$ and the length $m$ of an equilibrium-shaped $n$-sized fibril are not independent of each other, but interrelated by

$$i/m = \psi_h/\psi = a_h\sigma_h/a\sigma. \tag{8}$$

This important equation is nothing else, but a form of the Gibbs-Curie-Wulff theorem for the equilibrium shape of crystals (e.g., Refs. 61,62). It parallels that found by Kaischew[63] for rectangularly shaped 2D crystals. Equation 8 explains why the fibrils depicted in Fig. 2a of Ref. 31 become increasingly elongated with increasing the energy that characterizes the peptide bonding along the fibril $m$ axis. Indeed, since $\sigma$ or, equivalently, $\psi$ is greater when this energy is higher, Eq. (8) predicts a smaller fibril thickness-to-length ratio. For instance, for fibrils with $\psi = 10\psi_h$ or $20\psi_h$, the prediction is $i/m = 1/10$ or $1/20$, respectively.

We can now employ Eq. (5) to determine the dimensionless work $w_n \equiv W_n/kT$ to form an $n$-sized fibril with equilibrium shape [$W_n$ (J) is the dimensional work for formation of such a fibril]. Using $n = im$ and Eq. (8) yields

$$i = (\psi_h n/\psi)^{1/2}, \qquad m = (\psi n/\psi_h)^{1/2} \tag{9}$$

so that, upon eliminating $i$ and $m$ from Eq. (5) with the help of these relations, we obtain ($n \geq \psi/\psi_h$)

$$w_n = -sn + 4(\psi\psi_h)^{1/2}n^{1/2}. \tag{10}$$

This equation is of the form known from the classical theory of 2D nucleation (e.g., Ref. 56). Importantly, Eq. (10) is not valid for fibrils containing $n < \psi/\psi_h$ peptides and having the equilibrium shape, because according to Eq. (9) such small fibrils would be with the physically meaningless thickness of $i < 1$. This leads inescapably to the conclusion that the fibrils commence their ontogenesis as single β-sheets, i.e. as 1D aggregates, and continue it as 2D formations with equilibrium shape only after reaching the supersaturation-independent transition size $n_t$ given by

$$n_t = \psi/\psi_h = a\sigma/a_h\sigma_h. \tag{11}$$

For instance, fibrils characterized with $\psi/\psi_h = 10$ or $20$ will be single β-sheets until they contain $n \leq n_t = 10$ or $20$ peptides, respectively. The transition from 1D to 2D fibril geometry is clearly seen in Fig. 3 of Ref. 31.

Curves 1, 2 and 3 in Fig. 2 exhibit the dependence of the work $w_n$ to form a fibril of equilibrium shape on the fibril size $n$. They are drawn with the aid of Eq. (10) at $\psi/\psi_h = n_t = 10$ and scaled supersaturation $s/\psi_h = 1$, 2 or 3, respectively. The circles on the curves mark the fibril transition at $n = n_t$ from single β-sheet (1D aggregate) into fibril with equilibrium shape (2D aggregate). As seen from Fig. 2, beginning its growth as 1β-sheet, the fibril cannot keep growing in the same way when $n > n_t$, because this requires greater work (the dashed portions of straight lines 1β) than the work (curves 1, 2 and 3) for its growth with equilibrium shape. Thus, the full size dependence of the fibril formation work is given by Eq. (6) in the small-size range ($1 \le n \le n_t$) and by Eq. (10) when the fibril size is large enough ($n \ge n_t$). Figure 2 reveals also that at any $n > n_t$ the work $w_n$ from Eq. (10) is smaller not only than the work $w_{1,n}$ from Eq. (6), but also than the work

$$w_{i,n} = -(s - 2\psi_h / i)n + 2\psi i \qquad (12)$$

to form an $i$β-sheet of the same number $n$ of peptides and any given number $i = 1,2,3,...$ of β-sheets. Equation (12) parallels that for 1D formation of condensed phases (e.g., Ref. 56). It follows from Eq. (5) upon replacing $m$ by $n/i$, and at $i = 1$ it passes into Eq. (6). As visualized by the straight lines 2β and 3β in Fig. 2, given the $i$ value, Eq. (12) predicts a linear dependence of $w_{i,n}$ on $n$. The lines are drawn by using this equation with $\psi / \psi_h = n_t = 10$, $i = 2$ or 3 and scaled supersaturation $s / \psi_h = 1$, 2 or 3. It is seen that all these lines are above the curves depicting $w_n$ from Eq. (10) at the respective supersaturation. At a given $s$, successive portions of lines 1β, 2β and 3β form in fact a broken straight line enveloping the corresponding $w_n$ curve.

Figure 2 shows also that at a certain $s$ value the work $w_{2,n}$ does not change with $n$ (the uppermost line 2β has no slope). This means that at this supersaturation no work is done on attaching or detaching peptides to or from a 2β-sheet. Hence, regarded as a distinct peptide phase, the 2β-sheet is then in equilibrium or coexistence with the solution. Similarly, if conceived as different peptide phases, the 3β-, 4β-, etc. sheets can coexist with the solution, but at different supersaturations which we shall denote by $s_{i\beta}$. These coexistence supersaturations are readily obtained by setting equal to zero the bracketed factor in Eq. (12), because this factor is the driving force for $i$β-sheet growth or dissolution. Doing that leads to the formula ($i = 1,2,3,...$)

$$s_{i\beta} = 2\psi_h / i = 2a_h \sigma_h / ikT \qquad (13)$$

which shows that $s_{i\beta}$ decreases with increasing $i\beta$-sheet thickness $i$. At $i = 1$, i.e. for a single β-sheet, $s_{1\beta}$ assumes its maximum value $s_{1\beta} = 2\psi_h$ from Eq. (7). Recalling that $s = \Delta\mu / kT$ and employing Eqs. (A4) and (A9) in Eq. (13), we find that the equilibrium concentration $C_{i\beta}$ of peptide monomers and the equilibrium temperature $T_{i\beta}$ at which the $i\beta$-sheet neither grows nor dissolves at a given $T$ or $C_1$, respectively, are of the form ($i = 1,2,3,...$)

$$C_{i\beta} = C_e \exp(2a_h\sigma_h / ikT) \tag{14}$$

$$T_{i\beta} = T_e(1 - 2a_h\sigma_h / iL). \tag{15}$$

These expressions say that while $C_{i\beta}$ decreases, $T_{i\beta}$ increases with increasing the $i\beta$-sheet thickness $i$. This behavior of $C_{i\beta}$ and $T_{i\beta}$ is in agreement with that seen in a peptide phase diagram obtained by kinetic Monte Carlo simulations of aggregation of β-sheet forming peptides in solution.[55] In the limit of $i \to \infty$, i.e. when the $i\beta$-sheet is sufficiently thick, $C_{i\beta}$ and $T_{i\beta}$ are equal to the equilibrium concentration $C_e$ of peptide monomers and the equilibrium temperature $T_e$ that characterize the macroscopically large fibrillar phase.

## IV. NUCLEUS SIZE AND NUCLEATION WORK

Curve 1 in Fig. 2 depicts the change of $w_n$ with $n$ when fibril formation is nucleation-mediated and the fibrils have their equilibrium shape. It is seen that then $w_n$ passes through a maximum (marked by the star on curve 1) at a given fibril size $n*$. Thus, among the fibrils of different size, the $n*$-sized fibril is distinguishable with the greatest formation work. The fibril constituted of $n*$ peptides is the so-called nucleus (or critical nucleus), and the work $w* \equiv w_{n*} \equiv W_{n*} / kT$ for its formation is the nucleation work. This quantity is important, because it determines the energy barrier $(w* - w_{1,1})kT$ to nucleation of macroscopically large fibrils (the barrier magnitude at $s / \psi_h = 1$ is visualized by the double-headed arrow in Fig. 2). Treating $n$ as a continuous variable and employing the condition for maximum, $(dw_n / dn)_{n=n*} = 0$, from Eq. (10) we find that the CNT nucleus size and nucleation work depend on the supersaturation according to ($0 \le s \le s_{1\beta} = 2\psi_h$)

$$n* = 4\psi\psi_h / s^2 = 4aa_h\sigma\sigma_h / (kTs)^2 \tag{16}$$

$$w* = 4\psi\psi_h / s = 4aa_h\sigma\sigma_h / (kT)^2 s. \tag{17}$$

These equations parallel those obtained by Kaischew[63] for rectangularly shaped 2D crystals, and Eq. (16) is known as the Gibbs-Thomson equation (e.g., Ref. 56). We note that $n*$ and $w*$ are connected by the simple expression $w* = sn*$ and that they comply with the $\Delta\mu$ form[56,64,65] $dw*/ds = -n*$ of the nucleation theorem. As the nucleus thickness $i*$ and length $m*$ are related by Eq. (8) which characterizes the fibril equilibrium shape, from $n* = i* m*$ and Eq. (16) it follows that ($0 \le s \le s_{1\beta} = 2\psi_h$)

$$i* = 2\psi_h / s = 2a_h\sigma_h / kTs \tag{18}$$

$$m* = 2\psi / s = 2a\sigma / kTs. \tag{19}$$

The dashed lines in Fig. 3 display the supersaturation dependence of $i*$, $m*$, $n*$ and $w*$, predicted by the CNT Eqs. (16) – (19) at $\psi/\psi_h = n_t = 10$. As seen, all $i*$, $m*$, $n*$ and $w*$ decrease monotonically with increasing $s$. At the supersaturation $s_{1\beta} = 2\psi_h$ we have $i* = 1$, $m* = n* = \psi/\psi_h = n_t$ and $w* = w_{1,1}$ so that at this supersaturation the CNT nucleation barrier $(w* - w_{1,1})kT$ vanishes. This means that in the supersaturation range $s > s_{1\beta} = 2\psi_h$ CNT is not applicable. Indeed, then fibril formation occurs barrierlessly (see the connected solid lines 1β and 3 in Fig. 2) and, also, Eq. (18) yields the physically irrelevant result $i* < 1$. Thus, at the highest supersaturation $s_{1\beta}$ at which CNT is applicable, the CNT nucleus is a single β-sheet, $n_t$ peptides in length, formed without any work for the attachment of $n_t - 1$ peptides to the initial single peptide (see the middle line 1β in Fig. 2).

All of the above considerations are in the scope of CNT which provides a clear and mathematically simple description of the nucleation energetics. Recently, however, it has been shown[58] that this theory may underestimate considerably the nucleation work of crystals and, as a consequence, overestimate by many orders of magnitude the crystal nucleation rate. In the case of 2D nucleation, the reason for this grave inaccuracy is that CNT disregards the work $W_a$ (J) done on attaching the first molecule to the periphery of the crystal nucleus. This molecule triggers the propagation of a molecular row along the nucleus periphery and, thereby, the growth of the nucleus itself. We shall now apply CCNT[58] in order to quantitatively improve the CNT Eqs. (16) and (17) for the fibril nucleus size and nucleation work. As to the CNT Eq. (10) for $w_n$, CCNT does not provide a correction for it.

Figure 4 illustrates the CNT and CCNT fibril nuclei at one and the same supersaturation. The CNT nucleus (Fig. 4a) can preserve its equilibrium shape solely by growing in both thickness and length. However, as discussed also by Zhang and Muthukumar,[31] thickening is what actually impedes its growth because of the weak binding of the peptides to the surface of the nucleus outer β-sheets, i.e. to the nucleus $\sigma_h$-faces. The CNT nucleus needs just one more peptide (shown shaded in Fig. 4b) on one of these faces in order to be able to acquire barrierlessly the rest of the peptides necessary for the nucleus thickening by building-up of a new β-sheet. Indeed, while on attaching one peptide to the nucleus $\sigma_h$-face the work $-\Delta\mu = -skT$ is gained because of the peptide becoming part of the thermodynamically stable fibrillar phase, the work $2a\sigma = 2\psi kT$ is spent due to the augmenting of the nucleus total surface area by the area $2a$ of the peptide two $\sigma$-faces (as seen in Fig. 4, in this process the peptide two $\sigma_h$-faces do not contribute to the increase in the nucleus total surface area). Hence, the dimensionless overall peptide attachment work $w_a \equiv W_a / kT$ is

$$w_a = -s + 2\psi \ . \tag{20}$$

Subsequent attachment of a peptide to one of the $\sigma$-faces of the already attached first peptide does not change the nucleus total surface area so that in this process work is only gained, the gain being again $-\Delta\mu$. Similarly, further successive lateral attachment of peptides is thermodynamically favored, because it requires no work to be done until the complete building-up of a whole new β-sheet on the CNT nucleus.

It follows from the above that the CCNT fibril nucleus (Fig. 4b) is one peptide bigger than the CNT one (Fig. 4a) and that the CCNT nucleation work equals the CNT one plus the work for the first peptide attachment to a β-sheet surface. Thus, in view of Eqs. (16), (17) and (20), the CCNT formulae for the supersaturation dependence of the fibril nucleus size and the fibril dimensionless nucleation work read ( $0 \le s \le s_{1\beta} = 2\psi_h$ )

$$n^* = 4\psi\psi_h / s^2 + 1 = 4aa_h\sigma\sigma_h /(kTs)^2 + 1 \tag{21}$$

$$w^* = 4\psi\psi_h / s - s + 2\psi = 4aa_h\sigma\sigma_h /(kT)^2 s - s + 2a\sigma / kT \ . \tag{22}$$

These formulae are a central result of the present study. Their applicability is restricted to the CNT supersaturation range $[0, s_{1\beta}]$, and Eq. (21) is the corrected Gibbs-Thomson equation. In the $\psi_h = \psi$ case Eqs. (21) and (22) pass into the CCNT ones for 2D nucleation of square-shaped crystals.[58] It can be readily verified that, as it should be, $n*$ and $w*$ from Eqs. (21) and (22) obey the nucleation theorem in the form[56,64,65] $dw*/ds = -n*$. Also, it is seen that they are related by the simple formula $w* = (n*-2)s + 2\psi$.

The CCNT $n*(s)$ and $w*(s)$ dependences predicted by Eqs. (21) and (22) are represented in Fig. 3 by the solid lines n* and w*, respectively. We observe that while the CCNT correction has practically no effect on the CNT nucleus size, it affects strongly the CNT nucleation work. Importantly, while at $s = s_{1\beta}$ the CNT nucleation barrier vanishes, because at this supersaturation the CNT nucleation work $w*$ from Eq. (17) equals $w_{1,1}$, the CCNT nucleation barrier is still with the considerable height of $2(\psi - \psi_h)kT$ which is the difference between the CCNT $w*kT$ from Eq. (22) and $w_{1,1}kT$ at $s = s_{1\beta}$. CCNT thus predicts that fibril formation remains nucleation-mediated at supersaturations even higher than $s_{1\beta}$. However, at these high supersaturations fibril nucleation is non-classical, because considerations beyond the CNT requirement for equilibrium fibril shape are necessary for determination of the corresponding nucleation barrier and of the limiting supersaturation at which this barrier vanishes and above which fibril formation occurs barrierlessly, i.e. in the so-called metanucleation regime.[58]

## V. NUCLEATION RATE

In nucleation of fibrils in the volume of a supersaturated solution the nucleation rate $J$ ($m^{-3}$ $s^{-1}$) is the frequency of appearance of supernucleus fibrils per unit solution volume. According to the classical Szilard-Farkas model of nucleation (e.g., Refs. 56,57), random attachment and detachment of single peptides, i.e. peptide monomers, to and from a subnucleus fibril is the mechanism by which this fibril may eventually grow bigger than the nucleus and become a supernucleus. This mechanism of direct polymerization leads to a simple expression for the nucleation rate when the solution supersaturation and temperature are kept constant. Then nucleation occurs in stationary regime and its time-independent or stationary rate $J$ is given by the general formula (e.g., Refs. 56,57)

$$J = zf*C*. \tag{23}$$

Here $C^*$ (m$^{-3}$) is the equilibrium concentration of fibril nuclei in the solution and $f^*$ (s$^{-1}$) is the frequency of attachment of peptide monomers to a nucleus. The numerical parameter $z \leq 1$ is the so-called Zeldovich factor which takes into account that $C^*$ is about twice the stationary concentration of nuclei and that not every attachment event results in overgrowth of the nucleus to a macroscopic size. For 2D nucleation, according to CNT in its self-consistent formulation, $z$ and $C^*$ are expressed as (e.g., Ref. 56)

$$z = (s / 4\pi n^*)^{1/2} \tag{24}$$

$$C^* = C_1 \exp(w_{1,1} - w^*) \tag{25}$$

where $w_{1,1}$ is the dimensionless work for monomer formation (the monomer is formally considered as the smallest representative of the nucleating phase). Equation (25) is often used in the equivalent form $C^* = C_0 \exp(-w^*)$, because $C_1$, $w_{1,1}$ and the concentration $C_0$ (m$^{-3}$) of nucleation sites in the solution are related by the expression[56] $C_1 = C_0 \exp(-w_{1,1})$. As to $f^*$, it depends on the particular mechanism of monomer attachment to the nucleus. For instance, when the diffusion of monomer peptides toward the nucleus is fast enough, at the nucleus/solution interface there are always peptides ready to be incorporated into the nucleus by attaching themselves predominantly lengthwise, i.e. to the nucleus $\sigma$-faces. In this case, if $f_1$ (s$^{-1}$) is the frequency of lengthwise attachment of monomer peptides to a single β-sheet, $f^*$ is approximately given by

$$f^* = f_1 i^*, \tag{26}$$

because a nucleus with thickness of $i^*$ β-sheets has an $i^*$-fold greater chance for such an attachment than a single β-sheet. As it should be, at $i^* = 1$ Eq. (26) results in $f^* = f_1$. The attachment frequency $f_1$ itself is expected to be proportional to the concentration $C_1$ of monomer peptide in the solution and to the factor $\exp(-E_a / kT)$ in which $E_a$ (J) is the activation energy for lengthwise attachment of a peptide to a single β-sheet.

We can now obtain the CNT formula for the stationary rate $J$ of fibril nucleation. Substituting $z$, $C^*$ and $f^*$ from Eqs. (24) – (26) in Eq. (23), accounting that in accordance with Eq. (6) we have $w_{1,1} = -s + 2(\psi + \psi_h)$ and employing Eqs. (16) – (18), we arrive at the equation ($0 \leq s \leq s_{1\beta} = 2\psi_h$)

$$J = \left(\frac{\psi_h s}{4\pi\psi}\right)^{1/2} f_1(s)C_1 \exp\left[2(\psi + \psi_h) - s - \frac{4\psi\psi_h}{s}\right].$$ (27)

Using this equation should however be avoided, because it has been shown[58] that it is highly inaccurate for crystals with dimensionless specific surface energy $\psi > 1$. The quantitatively reliable formula for $J$ is the CCNT one and, to find it, we use again Eq. (23) with $C^*$ from Eq. (25), but with $w^*$ from the CCNT Eq. (22). In addition, we employ the approximation $z = 1/2$, because only half of the equilibrium nucleus concentration $C^*$ is effective in stationary nucleation[56] and because after attaching a peptide, the CCNT nucleus virtually always grows to a macroscopically large size. Also, in accordance with Eq. (26), we approximate $f^*$ by $f^* = 2f_1$, since in most cases the CCNT nucleus is expected to be two β-sheets thick (one β-sheet plus one peptide on the sheet). Thus, with the aid of these approximations for $z$ and $f^*$ and of the above expression for $w_{1,1}$, we obtain the CCNT $J(s)$ dependence in the form ($0 \leq s \leq s_{1\beta} = 2\psi_h$)

$$J = f_1(s)C_1 \exp\left[2\psi_h - \frac{4\psi\psi_h}{s}\right].$$ (28)

Comparing $J$ from Eqs. (27) and (28), we see that the CCNT nucleation rate is much lower that the CNT one, because the exponential function in Eq. (28) does not contain the summand $2\psi - s$ in which $\psi = a\sigma/kT$ is usually much greater than unity (see below). The absence of this summand cannot be compensated by the absence in Eq. (28) of the factor $(\psi_h s/4\pi\psi)^{1/2}$ which is typically a number between 0.01 and 0.1. Like Eq. (27), Eq. (28) is applicable in the CNT supersaturation range $[0, s_{1\beta}]$.

When $T$ is fixed and $s$ is controlled by means of the concentration $C_1$ of monomer peptide, we have $s = \ln(C_1/C_e)$ and $f_1 = f_{1,e}C_1/C_e$ (the attachment frequency $f_{1,e}$ ($s^{-1}$) is proportional to $\exp(-E_a/kT)$ and is the value of $f_1$ at the equilibrium concentration $C_e$ of monomer peptide, i.e. at $s = 0$). From Eq. (28) it then follows that the $J(C_1)$ dependence is of the form ($1 \leq C_1/C_e \leq \exp(2a_h\sigma_h/kT)$)

$$J(C_1) = A(C_1/C_e)^2 \exp[-B/\ln(C_1/C_e)],$$ (29)

where the $C_1$-independent kinetic factor $A$ ($m^{-3}$ $s^{-1}$) and the dimensionless thermodynamic parameter $B$ are given by

$$A = f_{1,e} C_e \exp(2a_h \sigma_h / kT) \tag{30}$$

$$B = 4aa_h \sigma \sigma_h / (kT)^2 . \tag{31}$$

Taking the logarithm of both sides of Eq. (29), differentiating with respect to $\ln C_1$ and accounting that the CCNT nucleus size $n*$ is given by Eq. (21), we find that

$$n*(C_1) = d(\ln J) / d(\ln C_1) - 1 . \tag{32}$$

This important formula shows that when available isothermal $J(C_1)$ data are plotted in $\ln J$ - vs.- $\ln C_1$ coordinates, the slope of the resulting line is a direct measure of the nucleus size. Equation (32) parallels that in Ref. 13 and, as can be readily verified, it holds true also for the CNT $n*$ and $J$ from Eqs. (16) and (27). This is so, because Eq. (32) is in fact a general result following from application of the nucleation theorem to isothermal nucleation of condensed single-component phases.[56,64,65] Given isothermal $J(C_1)$ data, Eq. (32) thus allows a theory-independent experimental determination of the nucleus size as a function of the concentration of monomer peptide in the solution and verification of any formula for the supersaturation dependence of $n*$, e.g., the corrected Gibbs-Thomson Eq. (21) of CCNT. Naturally, a more accurate determination of $n*$ requires replacement of $C_1$ in Eq. (32) by the peptide activity.

In the other case of experimental interest, the case when $s$ is controlled with the aid of $T$ at fixed $C_1$, the proportionality of $f_1$ to $\exp(-E_a / kT)$ can be expressed as $f_1 = f_0 \exp(-E_a / kT)$, where the virtually $T$-independent frequency factor $f_0$ ($s^{-1}$) is proportional to $C_1$. As then $s = L\Delta T / kT_e T$, Eq. (28) results in the following $J(T)$ dependence ($0 \le \Delta T \le 2a_h \sigma_h T_e / L$):

$$J(T) = A \exp(-E / kT) \exp(-B / T\Delta T) . \tag{33}$$

Here the kinetic factor $A$ ($m^{-3}$ $s^{-1}$), the thermodynamic parameter $B$ ($K^2$) and the effective activation energy $E$ (J) are specified by

$$A = f_0 C_1 \tag{34}$$

$$B = 4aa_h \sigma \sigma_h T_e / kL \tag{35}$$

$$E = E_a - 2a_h \sigma_h , \tag{36}$$

and it should be kept in mind that Eq. (33) is applicable provided the peptides remain in extended conformation in the entire temperature range studied (then all three parameters $A$, $B$ and $E$ in the equation can be treated as practically $T$-independent).

Equations (28), (29) and (33) are also a central result of the present study. The concentration dependence of $J$ from Eq. (29) is illustrated in Fig. 5 by the solid line. The line is drawn by using this equation with $a\sigma/kT = \psi = 12$ and $a_h\sigma_h = a\sigma/10$, the latter implying $a_h\sigma_h/kT = \psi_h = 1.2$. The value of $\psi$ is obtained with $T = 300$ K, $a = 5$ nm$^2$ and $\sigma = 10$ mJ/m$^2$, a specific surface energy in the range of 0.1–30 mJ/m$^2$ reported for protein crystals in aqueous solutions.[30,66-73] The $a$ value follows from $a = d_0 d$ with $d_0 = 5$ nm and $d = 1$ nm.[21,23]

As seen from Fig. 5, the monotonic increase of $J$ with $C_1$ is much faster at lower concentrations of monomer peptide than at those close to the concentration $C_{1\beta} = C_e \exp(2a_h\sigma_h/kT)$ corresponding to the maximal supersaturation $s_{1\beta} = \ln(C_{1\beta}/C_e)$ of CNT and CCNT applicability. This behavior of $J$ is due to the considerable diminishing of the nucleation work $w^*$ and the nucleus size $n^*$ with the supersaturation $s = \ln(C_1/C_e)$ approaching $s_{1\beta}$. The numbers at the circles on line CCNT in Fig. 5 indicate the CCNT $n^*$ values following from Eq. (21) at the corresponding supersaturations. For comparison between CCNT and CNT, the dashed line in Fig. 5 depicts the CNT isothermal nucleation rate $J$ from Eq. (27) (with $f_1 C_1 = f_{1,e} e^{2s}$, $s = \ln(C_1/C_e)$, $\psi = 12$ and $\psi_h = 1.2$), and the numbers at the circles on the line indicate the corresponding CNT nucleus size $n^*$ from Eq. (16). We observe that at the exemplified values of $\psi$ and $\psi_h$ the CNT nucleation rate is about 11 orders of magnitude higher than the CCNT one. This spectacular overestimation is almost entirely due to the CNT underestimation of the nucleation work $w^*$ (see Fig. 3), because the pre-exponential factors in Eqs. (27) and (28) differ by about one order of magnitude only. As to the $f_{1,e} C_e$ product which scales $J$ in Fig. 5, it may have values in a range which is orders-of-magnitude wide, because $f_{1,e}$ can hardly be higher than the attachment frequency of $10^5$ s$^{-1}$ encountered in nucleation of inorganic crystals in solutions (e.g., Refs. 56,57), but can easily be as low as the attachment frequency of $10^{-4}$ s$^{-1}$ inferred from measured rates of fibril elongation.[74] Hence, with exemplifying $C_e = 10^{20}$ m$^{-3}$ (i.e. 1.5 μM), it may be expected that in many cases of homogeneous nucleation the product $f_{1,e} C_e$ is from $10^{16}$ to $10^{25}$ m$^{-3}$ s$^{-1}$. This product can, however, be several orders of magnitude smaller for peptides with considerably inhibited solubility and/or ability of lengthwise attaching to β-sheets.

## VI. CONCLUSION

The analysis made shows that application of existing general theories of nucleation of new phases to amyloid fibril nucleation by the mechanism of direct polymerization can supply valuable information about both the thermodynamics and the kinetics of the process. The modeling of the nanosized amyloid fibrils (protofilaments) by prismatic aggregates with fixed width, changing thickness determined by the number of successively layered β-sheets in them, and changing length equal to the number of peptides in a β-sheet leads to the general CNT formula, Eq. (5), for the work to form a fibril of given size. This work is expressed by Eq. (12) when the fibril evolves with constant number of constituent β-sheets and by Eq. (10) when it preserves equilibrium shape during its evolution. Due to the impossibility of the smallest fibrils to be less than one β-sheet thick, a fibril first develops as an 1D aggregate and becomes a 2D one only later, after reaching the transition size $n_t$ given by Eq. (11). Both the transition size and the equilibrium shape of the bigger fibrils are controlled by the ratio of the specific surface energies $\sigma$ and $\sigma_h$ of the fibril faces normal to the fibril elongation and thickening axes. These energies affect also the fibril nucleus size $n^*$ and nucleation work $w^*$ whose CNT dependences on the supersaturation $s$ are given by Eqs. (16) and (17). The corresponding CCNT dependences are given by Eqs. (21) and (22) and have always to be used for a more accurate determination of $n^*$ and, especially, of $w^*$. Importantly, CCNT predicts that fibril formation remains nucleation-mediated at supersaturations even higher than the supersaturation $s_{1\beta}$ at which the CNT nucleation barrier vanishes. At these high supersaturations, however, fibril nucleation is non-classical, because the respective nucleation barrier cannot be determined with the help of the CNT requirement for equilibrium shape. Equations (21) and (22) show that changes in $\sigma$ and/or $\sigma_h$ brought about, e.g., by alteration of the solution ionic strength, adsorption of impurity molecules on the fibril faces or mutations along the peptide chain may cause significant changes in $n^*$ and $w^*$ and, thereby, in the fibril nucleation kinetics.

The kinetics of stationary fibril nucleation is characterized by the nucleation rate $J$ which is determined much more accurately by CCNT than by CNT because of the adequate calculation of the fibril nucleation work $w^*$. The general CCNT dependence of $J$ on the supersaturation $s$ is represented by Eq. (28), and the particular CCNT dependences of $J$ on the solution concentration $C_1$ or temperature $T$ are expressed by Eqs. (29) and (33). The nucleation rate is an exponentially strong function of $s$ because of the $s$-dependent barrier to fibril nucleation in the supersaturation range $[0, s_{1\beta}]$ of applicability of these equations. In isothermal nucleation, Eq. (32) allows a theory-independent determination of the number $n^*$ of peptides in the fibril nucleus from the slope of the line representing experimental $J(C_1)$ data in double logarithmic coordinates.

The results obtained are applicable to homogeneous amyloid fibril nucleation which can occur when the protein solution is sufficiently pure and/or strongly supersaturated. As in the classical theory of nucleation (e.g., Refs. 56,57), they need appropriate modification in order to be used in the case of heterogeneous nucleation which takes place on nucleation-active sites provided by foreign agents such as impurity nanoparticles within the solution (e.g., Refs. 26,75) and/or by foreign surfaces contacting the solution.

Currently, open questions in nucleation of amyloid fibrils are, e.g., the surface energies and the equilibrium shape of the fibrils, the size of the nucleus fibril, the magnitude of the nucleation barrier, the mechanism of peptide attachment to the fibrils and, most importantly, the dependence of the nucleation rate on the peptide concentration and solution temperature. The analysis made offers answers to some of these questions and the results obtained in the paper could be a helpful guide in studying the intriguing phenomenon of amyloid fibril nucleation.

## ACKNOWLEDGMENT

The work of one of the authors (S.A.) was supported by the EPSRC grant EP/G026165/1.

## APPENDIX A: THE SUPERSATURATION

In experiments on protein aggregation, the supersaturation $\Delta\mu$ is usually controlled either by the concentration of monomer protein at a fixed solution temperature (e.g., Refs. 9-14,16-18,29,30,33,36,38,44,52-54,68,70) or by the temperature at a fixed protein concentration (e.g., Refs. 9,10,14,19,25,26,70,74,76).

To express $\Delta\mu$ from Eq. (4) in terms of the concentration $C_1$ of monomer peptide at a fixed temperature $T$ we recall that when the solution is sufficiently dilute, thermodynamically, $\mu_s$ and $C_1$ are related by (e.g., Refs. 56,57)

$$\mu_s = \mu_r + kT \ln(C_1/C_r). \tag{A1}$$

Hence, $\Delta\mu$ from Eq. (4) becomes

$$\Delta\mu(C_1) = \mu_r + kT \ln(C_1/C_r) - \mu_f(C_1) - 2a_0\sigma_0(C_1). \tag{A2}$$

The $C_1$-independent reference chemical potential $\mu_r$, the reference concentration $C_r$ and the $\mu_f$ and $\sigma_0$ terms can be eliminated with the help of the equation

$$0 = \mu_r + kT \ln(C_e/C_r) - \mu_f(C_e) - 2a_0\sigma_0(C_e) \tag{A3}$$

which follows from Eq. (A2) upon taking into account that there is no driving force for fibril nucleation and growth when the solution is saturated [then $C_1$ equals the equilibrium concentration (or solubility) $C_e$ of monomer peptide at which, by definition, $\Delta\mu = 0$]. Since the chemical potential $\mu_f$ of the peptides in the bulk fibrillar phase is practically $C_1$-independent and the $\sigma_0(C_1)$ dependence may be negligible, by subtracting Eq. (A3) from Eq. (A2) and using the approximations $\mu_f(C_1) - \mu_f(C_e) \approx 0$ and $\sigma_0(C_1) - \sigma_0(C_e) \approx 0$, we obtain the known $\Delta\mu(C_1)$ formula for nucleation of condensed phases in solutions (e.g., Refs. 56,57)

$$\Delta\mu(C_1) = kT \ln(C_1/C_e). \tag{A4}$$

We note, however, that for more concentrated peptide solutions a more accurate evaluation of $\Delta\mu$ requires replacing the concentrations in Eq. (A4) by the corresponding activities.[56,57] Also, in some cases the neglected term $-2a_0[\sigma_0(C_1) - \sigma_0(C_e)]$ may need accounting for as a summand in the right-hand side of Eq. (A4).

In studying fibril nucleation and growth it is also important to know how $\Delta\mu$ varies with changing the solution temperature $T$ at a fixed monomer peptide concentration $C_1$. Then an approximate formula for the $\Delta\mu(T)$ dependence can be obtained from Eq. (4) with accounting for the thermodynamic relations

$$\mu_s = \mu_{r,s} - s_s T \tag{A5}$$

$$\mu_f = \mu_{r,f} - s_f T \tag{A6}$$

which are valid when the entropies $s_s$ and $s_f$ per peptide in, respectively, the solution and the bulk fibrillar peptide phase are practically $T$-independent in the temperature range studied. Combining Eqs. (4), (A5) and (A6) yields

$$\Delta\mu(T) = \mu_{r,s} - \mu_{r,f} - (s_s - s_f)T - 2a_0\sigma_0(T) \,. \tag{A7}$$

Similar to Eq. (A2), the $T$-independent reference chemical potentials $\mu_{r,s}$ and $\mu_{r,f}$ as well as the $\sigma_0$ term can be eliminated by using the equation

$$0 = \mu_{r,s} - \mu_{r,f} - (s_s - s_f)T_e - 2a_0\sigma_0(T_e) \tag{A8}$$

which expresses the fact that the solution is saturated ($\Delta\mu = 0$) at the equilibrium temperature $T_e$ (by definition, at $T = T_e$ the macroscopically large fibrillar phase coexists with the solution). Since to a first approximation $\sigma_0$ can be treated as $T$-independent, subtracting Eq. (A8) from Eq. (A7) and using the approximation $\sigma_0(T) - \sigma_0(T_e) \approx 0$ leads to the formula

$$\Delta\mu(T) = (L/T_e)\Delta T \tag{A9}$$

in which $L \equiv (s_s - s_f)T_e$ (in J) is the latent heat or enthalpy (per peptide) of fibril formation, and $\Delta T \equiv T_e - T$, when positive, is the experimentally controlled undercooling. We note that Eq. (A9) parallels the known $\Delta\mu(T)$ dependence for crystal nucleation in melts (e.g., Ref. 56). Naturally, a more accurate determination of $\Delta\mu$ requires allowing for the temperature dependence of the entropies $s_s$ and $s_f$ (Ref. 56) or, equivalently, of the latent heat $L$ of peptide fibrillation. Also, in some cases the neglected term $-2a_0[\sigma_0(T) - \sigma_0(T_e)]$ may need accounting for as a summand in the right-hand side of Eq. (A9).

Being expressed via quantities that refer to the equilibrium between two bulk phases, the solution and the macroscopically large fibrillar phase, the latent heat $L$ of fibril formation is a well-defined parameter independent of the fibril length and/or thickness. In both real and computer experiments $L$ can be determined by plotting solubility-vs-temperature data in $\ln C_e$-vs-$(1/T)$ coordinates.[31,54,55] If $L$ is $T$-independent, the resulting line is straight, and the $L$ value is obtainable from the slope of this line. Experiments on peptide solubility[54] reveal that $C_e$ may pass through a minimum at a certain temperature. The descending and ascending portions of the $C_e(T)$ line correspond to negative and positive $L$, respectively. Computer simulations[31,55] show that in the temperature range in which the peptide molecules in the solution are practically fully unfolded (stretched) and form β-sheets, $L$ is positive and practically $T$-independent. In this case, since nucleation and growth are possible only for $\Delta\mu > 0$, $\Delta T$ is positive and is called the undercooling (as done in the present paper). When protein aggregation is to occur in a temperature range in which $L$ is negative (e.g., Refs. 54,76), $\Delta T$ has to be negative in order to ensure that $\Delta\mu > 0$ and for that reason it then acquires the physical meaning of overheating. It is important to note also that knowing $C_e$ as a function of $T$ allows determining the $\Delta\mu(T)$ dependence more accurately than by Eq. (A9). Indeed, substitution of the $C_e(T)$ function in Eq. (A4) results in

$$\Delta\mu(T) = kT \ln[C_1 / C_e(T)] \tag{A10}$$

where now $C_1$ is fixed. Albeit implicitly, through the known dependence of $C_e$ on $T$, this equation automatically takes into account the temperature dependence and the sign of the latent heat $L$ of protein aggregation. As can be readily verified, Eq. (A10) simplifies to Eq. (A9) when $L$ is $T$-independent, because then $C_e$ obeys the integrated van't Hoff equation

$$C_e(T) = C_1 \exp\left[\frac{L}{k}\left(\frac{1}{T_e} - \frac{1}{T}\right)\right] \tag{A11}$$

which corresponds to integration condition $C_e(T_e) = C_1$. Naturally, the applicability of Eq. (A10) is restricted to sufficiently dilute protein solutions. For more concentrated solutions, $C_1$ and $C_e$ in Eq. (A10) should be replaced by the respective activities.

Finally, it is worth noting that, experimentally, factors as the solvent and the solution pH can also be used to control the protein aggregation and dissolution (e.g., Refs. 54,76). When these or other factors affect the equilibrium concentration $C_e$ of the protein monomers and the solution is sufficiently dilute, Eq. (A4) or Eq. (A10) allow determination of the solution supersaturation at the corresponding $C_1$ and $T$.

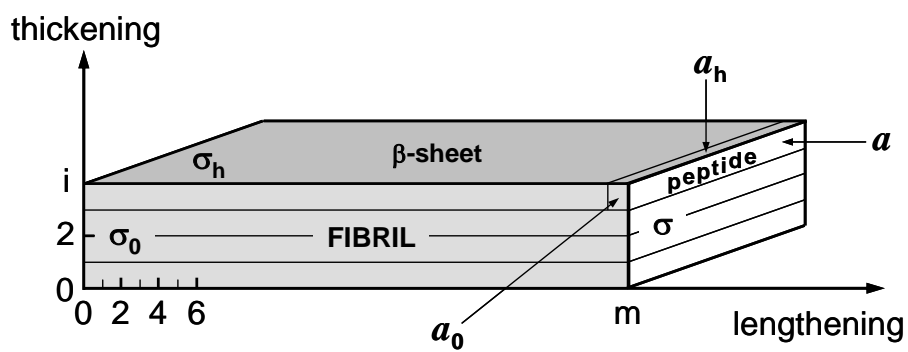

**Figure 1** Schematic of nanosized amyloid fibril (protofilament) with thickness of *i* β-sheets and length of *m* peptides. Only the uppermost of the four β-sheets in the fibril and the rightmost of the *m* consecutively aligned peptides in this β-sheet are labeled. The *σ*'s are the specific surface energies of the three fibril faces, and the *a*'s are the areas of the three peptide faces.

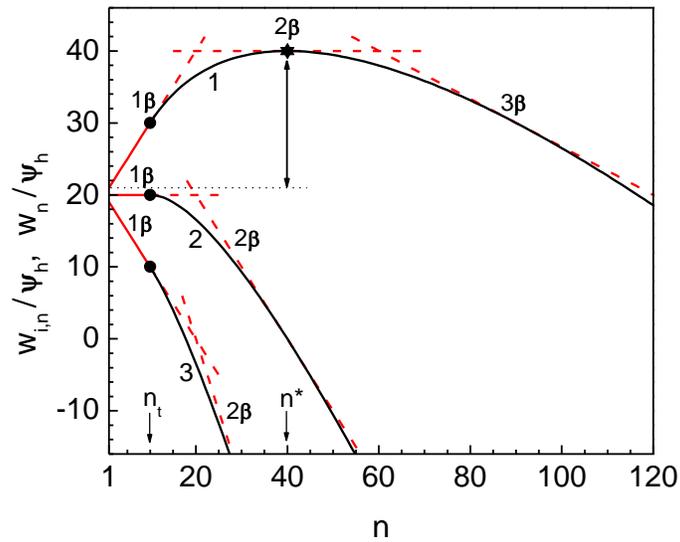

**Figure 2** Size dependence of the work $w_{i,n}$ to form 1β-sheet, 2β-sheet and 3β-sheet [according to the CNT Eq. (12), the straight lines] and the work $w_n$ to form a fibril with equilibrium shape [according to the CNT Eq. (10), the curves] at $\psi/\psi_h = 10$ and scaled supersaturation $s/\psi_h = 1$, 2 or 3 (as indicated). The double-headed arrow visualizes the height of the nucleation barrier at the nucleus size $n^*$, the circles on the solid lines correspond to the transition size $n_t$, and the star on line 1 indicates the maximum of $w_n$.

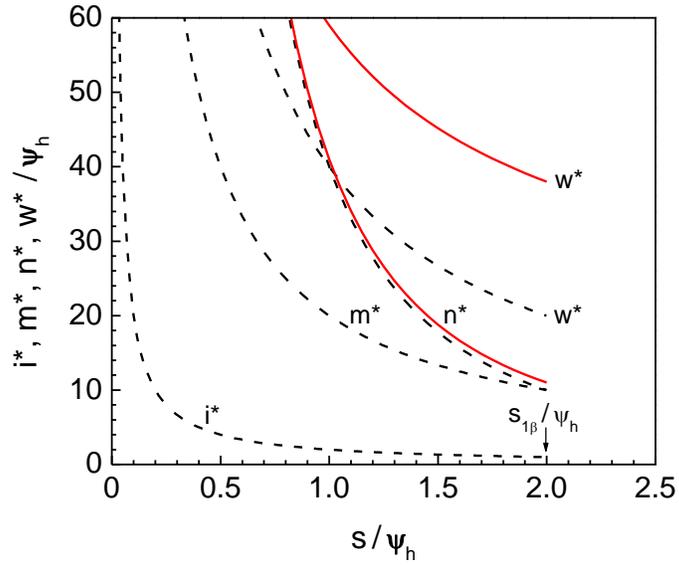

**Figure 3** Supersaturation dependence of the nucleation work $w^*$ and of the thickness $i^*$, length $m^*$ and size $n^*$ of the fibril nucleus at $\psi/\psi_h = 10$: solid lines – according to the CCNT Eqs. (21) and (22); dashed lines – according to the CNT Eqs. (16) – (19). The arrow indicates the maximum scaled supersaturation $s_{1\beta}/\psi_h$ of applicability of both CCNT and CNT.

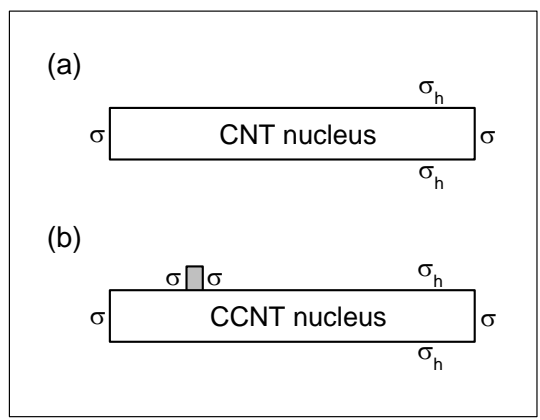

**Figure 4** Cross section of (a) CNT nucleus and (b) CCNT nucleus. The section is parallel to the *m,i* plane in Fig. 1, and the shaded rectangle schematizes the peptide that triggers the spreading of a new β-sheet on one of the nucleus $\sigma_h$-faces.

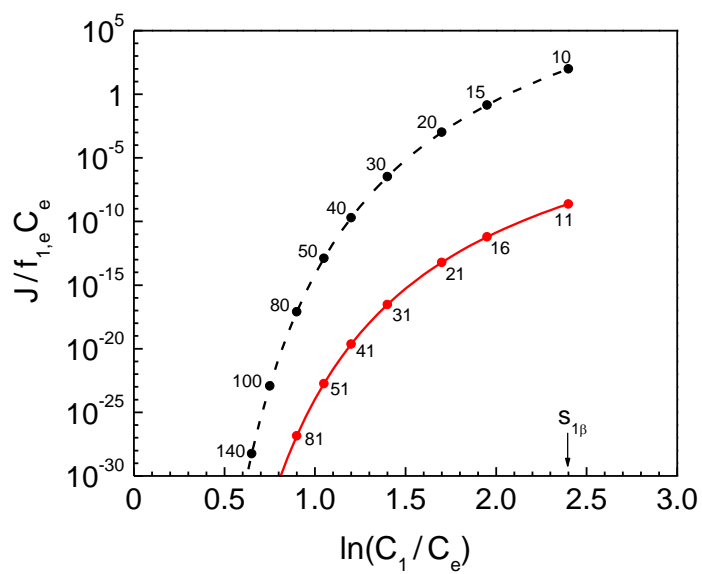

**Figure 5** Dependence of the fibril nucleation rate $J$ on the concentration $C_1$ of monomer peptide at $\psi = 12$ and $\psi / \psi_h = 10$: solid line – the CCNT Eq. (29); dashed line – the CNT Eq. (27). The arrow points to the maximum supersaturation $s_{1\beta}$ of applicability of both CCNT and CNT, and the numbers at the circles on the lines indicate the number of peptides in the CCNT or CNT fibril nucleus at the corresponding concentration.